%% file: main.tex
\documentclass[reprint,aps,physrev,amsmath,amssymb, nobalancelastpage, superscriptaddress]{revtex4-2}

\usepackage[utf8]{inputenc}
\usepackage[T1]{fontenc}
\usepackage{graphicx}
\usepackage{dcolumn}
\usepackage{bm}
\usepackage{hyperref}

\begin{document}

\title{Zero-Mean Oscillations Drive Recirculations in Potential Flows at Vanishing Reynolds Number}

\author{Alexandre S. Avaro}
\affiliation{Laboratoire d'Hydrodynamique (LadHyX), CNRS, École Polytechnique, Institut Polytechnique de Paris, 91120 Palaiseau, France
}

\author{Leon V. Gebhard}
\affiliation{Laboratoire d'Hydrodynamique (LadHyX), CNRS, École Polytechnique, Institut Polytechnique de Paris, 91120 Palaiseau, France
}

\author{Gabriel Amselem}
\email{gabriel.amselem@polytechnique.edu}
\affiliation{Laboratoire d'Hydrodynamique (LadHyX), CNRS, École Polytechnique, Institut Polytechnique de Paris, 91120 Palaiseau, France
}

\author{Charles N. Baroud}
\email{charles.baroud@ladhyx.polytechnique.fr} 
\affiliation{Laboratoire d'Hydrodynamique (LadHyX), CNRS, École Polytechnique, Institut Polytechnique de Paris, 91120 Palaiseau, France
}
\affiliation{Institut Pasteur, Université Paris Cité, Physical Microfluidics and Bioengineering, 25-28 Rue du Dr. Roux, 75015 Paris, France}

\begin{abstract}
Recirculating flows are central to transport phenomena, but their generation in potential flows requires additional physical mechanisms. We show experimentally that recirculation at vanishing Reynolds number can be produced at prescribed locations by coupling oscillations of a Hele-Shaw cell ceiling with an off-center passive obstacle. The net drift is well described by a potential flow model with no adjustable parameters. The resulting flows are localized around the obstacles and can be combined to generate programmable flow patterns at the microscale.
\end{abstract}

\maketitle

Recirculations are ubiquitous across scales and disciplines and play a critical role in material transport and trapping. Examples range from eddy currents in electric conductors~\cite{Kriezis1992PI} to vortices generated by jellyfish to capture prey~\cite{Dabiri2005JEB} to planetary-scale vortices such as Jupiter's Great Red Spot~\cite{Sommeria1988N}. Recirculations at large Reynolds numbers typically arise from inertia. At small Reynolds numbers, they emerge from the coupling of kinematic boundary conditions with the system geometry through viscous diffusion~\cite{Moffatt1964JFM, Jakiela2012PRL}.

In potential flow problems, which describe the motion of viscous fluid in porous media or the depth-averaged flow in the Hele-Shaw geometry, it is classically considered that recirculation is not possible~\cite{Panton2013}. In these flows, the velocity is proportional to the gradient of the potential. Therefore, the circulation must vanish around any closed contour, even in the presence of an obstacle. Although several studies have recently demonstrated the existence of recirculation in Hele-Shaw cells, they rely on additional physical mechanisms beyond those of potential flows~\cite{Gallaire2014PF, McKee2025PRF, Rallabandi2014JFM, Zhang2023JFM}. It remains unclear whether a closed recirculation can emerge from potential flows alone.

Here, we demonstrate with theory and experiments that a recirculation is possible by changing from an Eulerian to a Lagrangian framework. We consider a periodically deformed Hele-Shaw cell containing a fixed obstacle. The resulting oscillatory flow has zero mean and remains potential at all times. Yet, the obstacle rectifies it into a closed and localized Lagrangian recirculation that recalls ratchets in fluidic systems~\cite{Loutherback2009PRL, Kurzthaler2024JFM} and beyond~\cite{Norden2002APL}. We show that the resulting drift is purely kinematic, akin to Stokes drift in inertial traveling waves~\cite{Stokes1847TCPS, Longuet-Higgins1970JFM, vandenBremer2018PTRSA}, and that its strength is set by the signed area enclosed by the actuation cycle.

\begin{figure}[tb]
  \centering
  \includegraphics{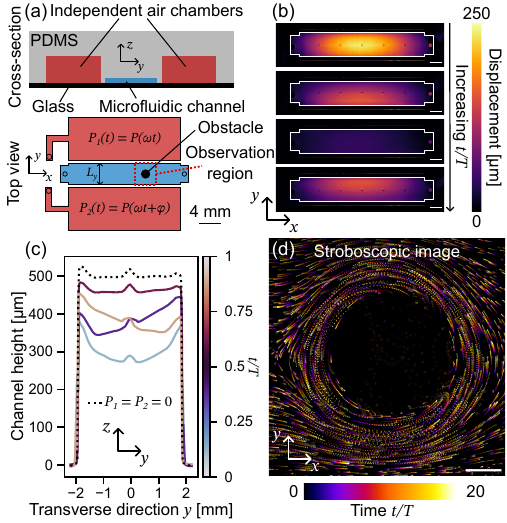}
  \caption{(a) Experimental setup: cross-section and top view of the microfluidic chip. (b) Channel ceiling deformation during one pressure actuation cycle, measured by fluorescence height profilometry in the absence of obstacle, for $\varphi = -\frac{\pi}{2}$. Scale bars: 2~mm. (c) Corresponding transverse height profiles at the center of the microfluidic channel. (d) Stroboscopic image of the flow in the microfluidic channel. Scale bar: 500~µm.}
  \label{fig1}
\end{figure}

The experimental setup consists of a PDMS (Sylgard 184) microfluidic chip comprising a channel of height $h_0=500$~µm, much smaller than its width $L_y=3.8$~mm and length $L_x = 20$~mm. The channel is flanked by two independent air chambers, of height $2.4$~mm and width $7.6$~mm, as shown in the schematic in Fig.~\ref{fig1}(a). Applying negative pressures to the air chambers via a multi-channel pressure controller (OB1 MK4, Elveflow) results in the deformation of the entire device. In particular, the ceiling of the microfluidic channel deflects downward, with a deflection amplitude proportional to the applied pressure~\cite{Gebhard2026,Jain2024B}.

Imposing different negative pressures in each air chamber yields an asymmetric channel ceiling deformation. We concentrate on periodic actuations. A pressure $P_1(t) = -\frac{P_o}{2} (1+\sin(\omega t))$ is imposed in one air chamber, and a pressure $P_2(t) = -\frac{P_o}{2} (1+\sin(\omega t + \varphi))$ in the other. Here, $P_o$ and $\omega$ respectively denote the amplitude and angular frequency of the actuation, and $\varphi$ the phase delay between the two actuations. The time evolution of the deformation of the entire microfluidic channel for $\varphi = -\frac{\pi}{2}$ and $P_o =500$~mbar is shown in Fig.~\ref{fig1}(b). The deformation profile is symmetric along $y$ when both actuating pressures are equal, and tilted otherwise, with a maximum deformation $\approx 250$~µm. Corresponding transverse channel height profiles along $y$ are shown in Fig.~\ref{fig1}(c). The tilt direction of the ceiling switches between the two halves of the pressure cycle, depending on the relative magnitudes of $P_1$ and $P_2$.

The microfluidic channel includes five small traps etched in the ceiling of the device and centered along $y$~\cite{Dangla2011PRL}. These traps appear as darker dots in Fig.~\ref{fig1}(b), and as a kink at $y=0$ in the profiles shown in Fig.~\ref{fig1}(c). A fixed obstacle, here a cylindrical hydrogel of radius $R$, is locally photopolymerized using a digital micromirror device (Primo, Alvéole) and centered at one of the traps. The hydrogel is composed of a mixture of 55\% polyethylene glycol (PEG, MW 200), 15\% polyethylene glycol diacrylate (PEGDA, MW 700), and 30\% water. Once polymerized, the obstacle adheres to the glass slide and remains at a fixed position even in the presence of actuation. The channel is then filled with PEG, a Newtonian fluid of kinematic viscosity $\nu=57~\text{mm}^2/\text{s}$~\cite{Sequeira2023JCED}, and seeded with fluorescent particles of diameter $3$~µm for flow visualization or diameter 500~nm for particle image velocimetry (PIV).

The actuation of the air chambers drives the fluid in and out of the channel through the left and right outlets. The resulting flow wraps around the obstacle and changes orientation and direction with the oscillating forcing. Imaging the channel once per actuation cycle reveals that the forcing generates a recirculating net flow localized around the obstacle, as shown in Fig.~\ref{fig1}(d) for $\varphi = -\frac{\pi}{2}$ (see also Movie~S1 in~\cite{SM}). This stroboscopic image is effectively a Poincaré map: it does not show the full tracer trajectories, but only captures the net flow over multiple cycles. Note that a drift occurs even in the absence of the obstacle due to the spatial variations of the ceiling height. Here, we concentrate on the localized recirculation that results from the presence of the obstacle.

\begin{figure*}[tb]
  \centering
  \includegraphics{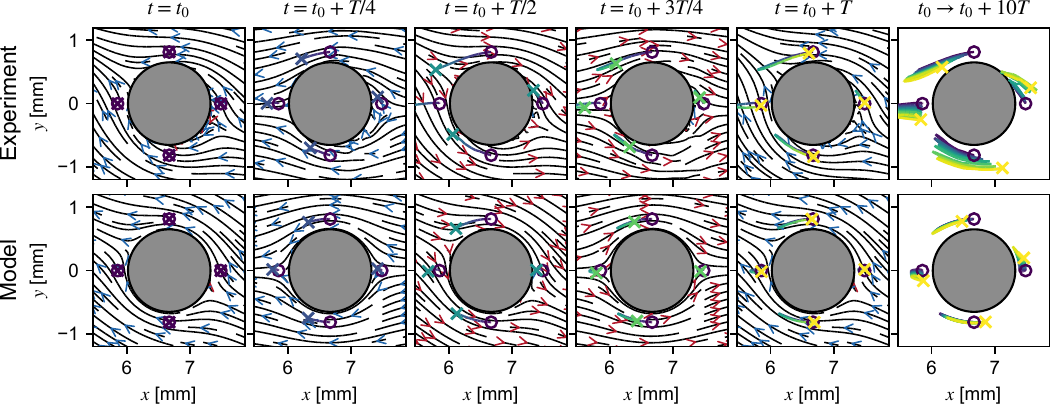}
  \caption{Experimental and predicted streamlines around the obstacle. Top row: streamlines deduced from instantaneous PIV measurements between two successive frames at five time instants over one period. Bottom row: predicted streamlines from the potential flow model at the same time instants. Red and blue arrows respectively indicate positive and negative horizontal velocity. The trajectories of four virtual tracers are indicated on all panels (initial positions are indicated with open circles, and current positions with crosses). The rightmost panels show the trajectories over 10 periods.}
  \label{fig2}
\end{figure*}

To understand the origin of the net recirculation, the flow profile around the obstacle is reconstructed at each instant using PIV~\cite{Thielicke2014JORS}. The experimental streamlines are shown at five instants in time in the top row of Fig.~\ref{fig2}. Near the obstacle, the flow resembles a uniform flow past a cylinder whose direction rotates due to the phase delay between the actuation sources. The rotation of the streamlines around the obstacle points to the mechanism responsible for the net drift: the flow is zero-mean and time-periodic, but passive tracers experience varying velocities along their trajectories, which leads to an overall displacement per actuation period. This can be seen in Fig.~\ref{fig2}, where virtual tracer trajectories show a small but systematic offset between their initial (circles) and final (crosses) positions after one period.

The net recirculation can be modeled using potential flow theory to describe the flow in the chip, treated as a Hele-Shaw cell~\cite{McKee2025PRF,Milne-Thomson1960}. Inertial effects are indeed negligible in these experiments, as both the Reynolds number $\mathrm{Re}=\frac{U_{i}h_0}{\nu}\approx10^{-3}$ (where $U_{i}$ is a characteristic instantaneous velocity) and the Womersley number $\mathrm{Wo}^2=\frac{\omega h_0^2}{\nu}\approx10^{-2}$ are very small. Consider first a channel without any obstacle, of dimensions $L_x$ and $L_y$. The Eulerian flow field satisfies $\mathbf{u}_E = \nabla \phi_0$, where the velocity potential $\phi_0$ obeys the Poisson equation whose source term represents the effects of vertical ceiling actuation throughout the domain:
\begin{equation}
  \Delta \phi_0 = \frac{1}{h_0} \frac{\partial \delta}{\partial t}.
  \label{poisson}
\end{equation}
Here, $\delta$ is the ceiling displacement. Experimentally, the ceiling deformation is not constant along $x$, but shows a maximum in the center of the device ($x=0$) and a minimum at the device edges ($x = \pm L_x/2$), which is modeled as a cosine profile $\cos\left(\pi x/L_x \right)$ (Fig.~\ref{fig1}(b)). Further, we approximate $\delta$ as a linear function of $y$ (Fig.~\ref{fig1}(c)). $\delta$ is therefore given by:
\begin{equation}
  \delta(x,y,t) = \left[\frac{\delta_1 + \delta_2}{2} + \frac{\delta_1 - \delta_2}{L_y}y \right] \cos(kx),
\end{equation}
where $\delta_1(t)$ and $\delta_2(t)$ correspond to the maximal deformations of the ceiling at $\pm L_y/2$, and $k=\pi/L_x$. Since the channel ceiling deformation scales linearly with the applied pressure~\cite{Gebhard2026}, $\delta_1$ and $\delta_2$ are respectively assumed to be proportional to $P_1$ and $P_2$. The boundary conditions are chosen to enforce impermeability at the side walls ($y=\pm L_y/2$), and identical pressure values on the channel ends ($x=\pm L_x/2$). The potential $\phi_0$ for the flow in the absence of obstacle can be derived analytically by solving Eq.~(\ref{poisson})~\cite{SM}. 

Assuming the obstacle is small compared to the domain ($kR \ll 1$), the velocity can be approximated as locally uniform on the scale of the obstacle~\cite{SM}. The flow profile is then estimated using $(U,V)=\nabla \phi_0(x_c,y_c)$, where $(x_c,y_c)$ is the obstacle center. The perturbed complex potential $\Phi_1$, and thus the flow around the obstacle, follows from the Milne-Thomson circle theorem~\cite{SM, Milne-Thomson1960}. The streamlines predicted using the results of the potential model are plotted in the bottom row of Fig.~\ref{fig2}. Theoretical streamlines complete a full rotation around the obstacle over one period, and are in excellent agreement with experimental results. 

We can now compute the trajectories of passive tracers in the flow. The time-varying Lagrangian velocity $\mathbf{u}_L(\mathbf{x}_0,t)$ of a tracer with initial position $\mathbf{x}_0$ in this unsteady flow is related to the local Eulerian velocity $\mathbf{u}_E(\mathbf{x},t)$ through~\cite{Stokes1847TCPS}:
\begin{equation}
  \mathbf{u}_L(\mathbf{x}_0,t) = \mathbf{u}_E\left(\mathbf{x}_0+\int_0^t\mathbf{u}_L(\mathbf{x}_0,t')dt',~t\right).
\end{equation}
At first order in tracer excursion, the drift given by the potential flow solution vanishes over a period $T$, i.e., $\langle \mathbf{u}_L\rangle_T \approx \langle\mathbf{u}_E \rangle_T=0$~\cite{SM}. At second order, a nonzero drift can be estimated~\cite{SM}:
\begin{equation}
  \langle \mathbf{u}_L\rangle_T \approx \frac{2R^4}{r^5} \frac{\sin(2kx_c)}{k^3h_0^2L_y} \left[ 1-\frac{\cosh(ky_c)}{\cosh(kL_y/2)} \right] \frac{\mathcal{A}}{T} \ \mathbf{e}_\theta,
\end{equation}
where $r$ is the distance to the center of the obstacle, and $\mathcal{A}$ denotes the signed area enclosed by one cycle of the actuation loop in the ($\delta_1$, $\delta_2$) plane. As observed in the Poincaré map of Fig.~\ref{fig1}(d), the net drift is purely azimuthal. Interestingly, this expression decouples the role of the actuation from the role of the obstacle and particle positions in the channel. Note for instance that two obstacles placed at streamwise positions symmetric with respect to the center of the chip will induce counter-rotating recirculations of equal intensity. In the specific case of sinusoidal actuation proportional to the profiles of $P_1$ and $P_2$ mentioned above~\cite{SM}:
\begin{equation}
\langle \mathbf{u}_L\rangle_T \approx \frac{\alpha^2 R^4}{4r^5} \frac{\sin(2kx_c)}{k^3h_0^2L_y} \left[ \frac{\cosh(ky_c)}{\cosh(kL_y/2)} -1\right] \omega P_o^2 \sin \varphi \ \mathbf{e}_\theta,
\label{sindrift}
\end{equation}
where the constant $\alpha$ is defined such that $\alpha P_i=-\delta_i$. Fluorescence height profilometry measurements yield $\alpha=0.5$~µm/mbar. The net drift accumulates over successive cycles, as shown in the rightmost panels of Fig.~\ref{fig2}. 

\begin{figure}[tb]
  \centering
  \includegraphics{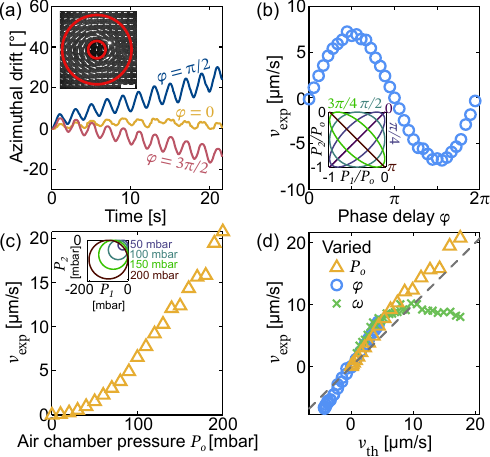}
  \caption{(a) Azimuthal drift of particles around the obstacle. Inset: PIV results on an example Poincaré map. Red circles indicate the boundaries of the annular integration region (scale bar: 500~µm). (b) Average azimuthal velocity versus phase delay $\varphi$. Inset: Actuation loops in the ($P_1/P_o$, $P_2/P_o$) plane for varying $\varphi$. (c) Average azimuthal velocity versus air chamber pressure $P_o$. Inset: Actuation loops in the ($P_1$, $P_2$) plane for varying $P_o$. (d) Experimental versus theoretical average azimuthal velocity for 86 measurements, each one averaged over 30 periods. Unless varied, the experimental parameters were the following: $\varphi=\frac{\pi}{2}$, $\omega = \pi~\mathrm{rad/s}$, $x_c = 3.33$~mm, $y_c = 0$~mm, $R=375$~µm, and $P_o=100$~mbar. The gray dashed line indicates $v_{\mathrm{th}} = v_{\mathrm{exp}}$.}
  \label{fig3}
\end{figure}

The angular displacement of fluid particles around the obstacle is determined experimentally using the computed trajectories of virtual particles seeded in an annulus of outer radius $d=4R$ around the obstacle, as shown in the inset of Fig.~\ref{fig3}(a). The angular displacement oscillates with a net overall drift that depends on the phase delay $\varphi$. The oscillations reflect the sinusoidal pressure actuation at period $T = 2$~s. When the actuation is in phase ($\varphi = 0$), there is no net angular drift. When the actuation is asymmetric ($\varphi = \pi/2$ or $3\pi/2$), the drift is nonzero and its direction depends on the value of the phase delay $\varphi$, as shown in Fig.~\ref{fig3}(a).

We then compute the experimental average (clockwise) azimuthal drift velocity $v_{\mathrm{exp}}$ in the same annulus for different values of the phase delay $0\leq \varphi\leq 2\pi$, actuation frequency $0\leq\omega\leq 4\pi \; \mathrm{rad/s}$, and actuation amplitude $0\leq P_o\leq 200$~mbar, averaged over 30 periods. The experimental velocity shows a sinusoidal dependence on $\varphi$ (Fig.~\ref{fig3}(b)), a linear dependence on the actuation frequency $\omega$ (see Fig.~S1 in~\cite{SM}), and a quadratic dependence on the actuation amplitude $P_o$ (Fig.~\ref{fig3}(c)), in agreement with the theoretical scalings of Eq.~(\ref{sindrift}). 

To compare experimental and theoretical results, the theoretical average azimuthal velocity $v_{\mathrm{th}}$ is calculated in the same annulus around the obstacle using Eq.~(\ref{sindrift}). The plot of $v_{\mathrm{exp}}$ as a function of $v_{\mathrm{th}}$ shows that all data points collapse onto a single line (Fig.~\ref{fig3}(d)). This agreement is remarkable given that the model contains no adjustable parameters, and that the boundary conditions for the channel deformation represent a simplification of the actuation mechanism over the whole channel. Experiments and theory disagree most at frequencies $\omega> 2\pi~\mathrm{rad/s}$, which we attribute to the finite response time of the pneumatic control system to the actuating pressure.

\begin{figure}[tb]
  \centering
  \includegraphics{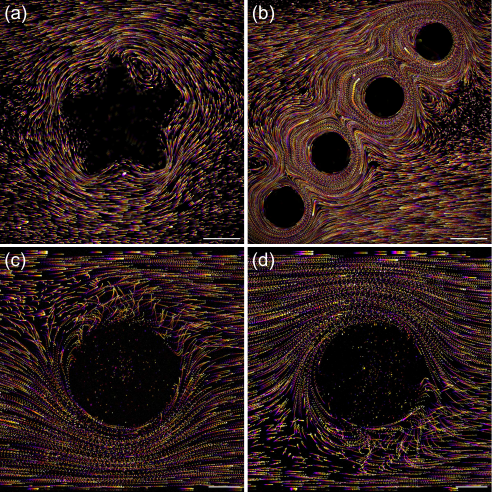}
   \caption{Combination of patterned flow fields. (a) Poincaré map for the flow around a five-pointed star obstacle. (b) Poincaré map for the flow field resulting from the arrangement of four neighboring obstacles positioned diagonally with respect to the streamwise direction of the channel. (c), (d) Poincaré maps for the superposition of an externally controlled Poiseuille flow and the recirculation induced by the obstacle and the deformable ceiling, for $\varphi=-\frac{\pi}{2}$~(c) or $\varphi=\frac{\pi}{2}$~(d). Scale bars: 500~µm.}
  \label{fig4}
\end{figure}

Armed with this quantitative understanding of the recirculation, we now exploit this framework to create complex flow patterns in the device. Changing the shape of the obstacle modifies the recirculating pattern: for instance, a star-shaped obstacle still leads to a recirculation, albeit with a more complex pattern (Fig.~\ref{fig4}(a)). Similarly, combining multiple obstacles within the same channel leads to multiple recirculating patterns, which interact if the obstacles are close enough. Obstacles aligned along a diagonal line across the channel yield net transport along the axis defined by the arrangement of the obstacles (Fig.~\ref{fig4}(b), see also Movie~S2 in~\cite{SM}).

Complex flow patterns can likewise emerge from combining actuation with an externally imposed flow, yielding a local flow router. We impose a constant flow rate $Q = 1~\text{µL/min}$ in the microfluidic channel using a syringe pump (Pump 11 Elite, Harvard Apparatus). Without pneumatic actuation, the flow goes from left to right everywhere in the channel. For actuation at $\varphi = -\frac{\pi}{2}$, $\omega = \pi~\mathrm{rad/s}$, $P_o = 100$~mbar, the flow remains directed from left to right below the obstacle, but it is canceled by the recirculating flow above it (Fig.~\ref{fig4}(c), see Movie~S3 in~\cite{SM}). When the phase delay is reversed to $\varphi=\frac{\pi}{2}$, the asymmetry is inverted (Fig.~\ref{fig4}(d), see Movie~S3 in~\cite{SM}). The fluid can thus be directed selectively to either side of the channel depending on the sign of $\sin \varphi$.

In summary, the results above show that a static, passive obstacle in an oscillating Hele-Shaw cell rectifies a global, periodic wall deformation into a steady Lagrangian recirculation confined to its neighborhood. The net drift is due to purely kinematic effects, analogous to the well-known Stokes drift, and is well described by a potential flow model. This result recalls previous theoretical studies that predicted that periodically reoriented potential flows can serve to control Lagrangian trajectories and promote chaotic advection~\cite{Lester2009PRE}. However, we now show that the presence of an obstacle in the domain confines the drift to its vicinity. This localization emerges from the strong local velocity gradients around the obstacle, which interact with the oscillating Eulerian field to generate the net drift through a second-order effect. 

The model developed above shows that the displacement per cycle is set by a single geometric quantity: the signed area enclosed by the actuation cycle. A loop of zero area (e.g., $\varphi = 0~\text{or}~\pi$, a straight segment) produces no drift, which recalls the scallop theorem~\cite{Purcell1977AJP, Shapere1989JFM, Amselem2023PRA}. Varying $\varphi$ and $P_o$ (Figs.~\ref{fig3}(b) and \ref{fig3}(c)) are therefore two different ways of varying the area $\mathcal{A}$ enclosed by these loops, as shown in the insets of the corresponding panels. This dependence also predicts that the drift amplitude is independent of the shape of the waveform. 

The flow structures constructed using this principle can be positioned, shaped, and combined either together or with an imposed external flow, making the topology of transport an accessible design variable at vanishing Reynolds number. Natural extensions of this work include mobile obstacles or additional actuators, which would offer further degrees of freedom for programmable flow control. Beyond the current experimental setup, the rectification of a periodically reoriented potential flow in the vicinity of an obstacle is a kinematic mechanism that does not depend on the underlying physical nature of the potential. It could therefore be extended to other transport problems modeled using the same equations, including flows in porous media or oscillating electric fields~\cite{Box2020PNAS, McKee2025PRF}. 

\begin{acknowledgments}
The authors thank the X-Fab of École Polytechnique for its help and fruitful discussions. This work was funded by the European Union (ERC-2023-ADG grant number 101142018 MELCART). During the preparation of this work, the authors used Claude Opus 5 for code generation, assistance with model derivation, and improvement of the readability of the text. The authors reviewed and edited the content as needed and take full responsibility for the content of the publication.
\end{acknowledgments}

\bibliography{biblio}

\end{document}


\title{Supplemental Material for Zero-Mean Oscillations Drive Recirculations in Potential Flows at Vanishing Reynolds Number}

\author{Alexandre S. Avaro}
\affiliation{Laboratoire d'Hydrodynamique (LadHyX), CNRS, École Polytechnique,
 Institut Polytechnique de Paris, 91120 Palaiseau, France}
\author{Leon V. Gebhard}
\affiliation{Laboratoire d'Hydrodynamique (LadHyX), CNRS, École Polytechnique,
 Institut Polytechnique de Paris, 91120 Palaiseau, France}
\author{Gabriel Amselem}
\email{gabriel.amselem@polytechnique.edu}
\affiliation{Laboratoire d'Hydrodynamique (LadHyX), CNRS, École Polytechnique,
 Institut Polytechnique de Paris, 91120 Palaiseau, France}
\author{Charles N. Baroud}
\email{charles.baroud@ladhyx.polytechnique.fr}
\affiliation{Laboratoire d'Hydrodynamique (LadHyX), CNRS, École Polytechnique,
 Institut Polytechnique de Paris, 91120 Palaiseau, France}
\affiliation{Institut Pasteur, Université Paris Cité, Physical Microfluidics and Bioengineering, 25-28 Rue du Dr. Roux, 75015 Paris, France}
 
\maketitle
 
\section{Supplemental Movies}
\noindent\textbf{Movie S1. Flow around an example passive obstacle.} The time-resolved frames show the instantaneous oscillatory flow. The stroboscopic sequence (i.e., sampled once per actuation cycle) reveals the net recirculating drift.

\noindent\textbf{Movie S2. Flow around four obstacles arranged diagonally.} The stroboscopic movie shows tracers both recirculating around each obstacle and drifting along the axis of the array.

\noindent\textbf{Movie S3. Recirculation combined with an imposed pressure-driven flow.} The stroboscopic movie shows the resulting drift under simultaneous air chamber actuation and pressure-driven flow applied using a syringe pump connected to the inlet of the channel. Reversing the actuation phase delay $\varphi$ reverses the recirculation, moving the region of the domain where the net horizontal velocity vanishes.

\section{Derivation of the drift expression}

\input{poisson_derivation}

\section{Velocity measurements for varying actuation frequency}
\begin{figure}[ht]
  \centering
  \includegraphics{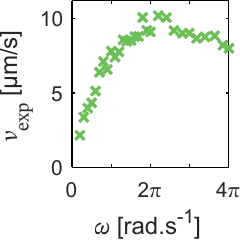}
  \caption{Average azimuthal drift velocity $v_{\mathrm{exp}}$ versus
  actuation frequency $\omega$. Other experimental parameters are identical to those reported for Fig.~3 of the main text.}
  \label{figs1}
\end{figure}

%% file: poisson_derivation.tex
\subsection{Derivation of the potential in the absence of obstacle}
We first derive the potential solution $\phi_0$ in the absence of obstacle. We consider a Hele-Shaw cell with dimensions $L_x$ and $L_y$, such that $x\in[-L_x/2, L_x/2]$ and $y\in[-L_y/2, L_y/2]$. Depth-averaged mass conservation yields:
\begin{equation}
    \frac{\partial h}{\partial t} + \nabla \cdot(h\mathbf{\langle u \rangle}_z) = 0,
\end{equation}
where $h(x,y,t)$ is the local height of the channel and $\mathbf{\langle u \rangle}_z$ is the depth-averaged velocity field. We define the ceiling deformation $\delta$ such that $h(x,y,t) = h_0 - \delta(x,y,t)$. Assuming $\delta/h_0 \ll 1$, the previous equation yields:
\begin{equation}
    \nabla \cdot\mathbf{\langle u \rangle}_z = \frac{1}{h_0} \frac{\partial \delta}{\partial t}.
\end{equation}
In the Hele-Shaw cell we can define a potential $\phi_0$ such that $\mathbf{\langle u \rangle}_z = \nabla \phi_0$. This yields the following Poisson equation:
\begin{equation}
    \Delta \phi_0 = \frac{1}{h_0} \frac{\partial \delta}{\partial t}.
\end{equation}
From the shape of the deformation shown in Fig. 1 of the main text, we assume that the deformation profile is the following:
\begin{equation}
    \delta(x,y,t) = \left[\frac{\delta_1 + \delta_2}{2} + \frac{\delta_1 - \delta_2}{L_y}y \right] \cos(kx),
\end{equation}
where $k = \pi/L_x$. This corresponds to a linear deformation in $y$. Here, $\delta_1$ and $\delta_2$, once weighted in $x$ by $\cos(kx)$, correspond to the deformations on either side of the channel. We assume that $\delta_i$ is proportional to $P_i$.

The problem then becomes:
\begin{equation}
    \Delta \phi_0 = [\Sigma + \Gamma y]\cos(kx),
\end{equation}
with
\begin{equation}
    \Sigma = \frac{\dot \delta_1 + \dot \delta_2}{2h_0},
\end{equation}
\begin{equation}
    \Gamma = \frac{\dot \delta_1 - \dot \delta_2}{h_0 L_y}.
\end{equation}
We use the following boundary conditions:
\begin{equation}
    \frac{\partial\phi_0}{\partial y} \left(x, y=\pm \frac{L_y}{2},t \right) = 0,
\end{equation}
\begin{equation}
    \phi_0 \left(x=\pm \frac{L_x}{2}, y,t \right) = 0.
\end{equation}
They respectively correspond to no flux through the side walls, and identical pressure at both ends of the channel. Using separation of variables, we look for solutions of the form $\phi_0 = f(y)\cos(kx)$ (compatible with the boundary conditions at $\pm L_x/2$). The problem then becomes the ordinary differential equation:
\begin{equation}
    f'' - k^2 f = \Sigma +\Gamma y.
\end{equation}
A particular solution is $f_p =-\frac{\Sigma+\Gamma y}{k^2}$. The homogeneous solution is $f_h = C_1 \cosh(ky) + C_2 \sinh(ky)$. The boundary conditions at $\pm L_y/2$ yield:
\begin{equation}
    -\frac{\Gamma}{k^2} + C_1k \sinh\left(\frac{kL_y}{2}\right) + C_2k \cosh\left(\frac{kL_y}{2}\right) = 0,
\end{equation}
\begin{equation}
    -\frac{\Gamma}{k^2} - C_1k \sinh\left(\frac{kL_y}{2}\right) + C_2k \cosh\left(\frac{kL_y}{2}\right) = 0.
\end{equation}
These conditions entail $C_1 = 0$ and $C_2 = \frac{\Gamma}{k^3 \cosh(kL_y/2)}$. The potential solution in the absence of the obstacle is therefore:
\begin{equation}
    \phi_0 (x,y,t)= \left(\frac{\Gamma}{k^3} \frac{\sinh(ky)}{\cosh(kL_y/2)} - \frac{\Sigma+\Gamma y}{k^2} \right) \cos(kx).
\end{equation}

\subsection{Perturbation description of the obstacle}
We now treat the obstacle as a perturbation to the potential field. If the obstacle is smaller than the characteristic length of $\phi_0$ (i.e., $kR \ll 1$), we can approximate the flow as locally uniform. We define the velocities at the center of the obstacle $(x_c,y_c)$: 
\begin{equation}
    U(t) = \frac{\partial\phi_0}{\partial x} (x_c,y_c,t) = \frac{\sin(kx_c)}{k} \left[\Sigma + \Gamma G(y_c) \right],
\end{equation}
\begin{equation}
    V(t) = \frac{\partial\phi_0}{\partial y} (x_c,y_c,t) = \cos(kx_c) H(y_c)\Gamma,
\end{equation}
where:
\begin{equation}
    G(y_c) = y_c - \frac{\sinh(ky_c)}{k \cosh(kL_y/2)}
\end{equation}
and
\begin{equation}
    H(y_c) =\frac{1}{k^2} \left[\frac{\cosh(ky_c)}{\cosh(kL_y/2)} - 1\right].
\end{equation}
The complex potential $\Phi_1(z)$ in the presence of the obstacle can be derived using the Milne-Thomson circle theorem:
\begin{equation}
    \Phi_1(z) = \overline{\mathcal{U}} z + \mathcal{U} \frac{R^2}{z},
\end{equation}
where $\mathcal{U} = U + iV$, $R$ is the radius of the obstacle, and $z$ the complex coordinate whose origin lies at the center of the obstacle. $\overline{\cdot}$ denotes the complex conjugate. It follows that:
\begin{equation}
    u - iv = \overline{\mathcal{U}} - \mathcal{U} \frac{R^2}{z^2},
    \label{cylinderflow}
\end{equation}
where $u$ and $v$ are the $x$ and $y$ components of the Eulerian velocity vector $\mathbf{u}_E(\mathbf{x},t)$. 

\subsection{Mathematical interlude}

We first define the following operations given a vector $\mathbf{n}$:

\begin{equation}
    \tilde{\mathbf{n}}(t) = \int_0^t \mathbf{n}(t')dt',
\end{equation}
\begin{equation}
    \dot{\mathbf{n}} = \frac{d \mathbf{n}}{dt},
\end{equation}
\begin{equation}
    \langle\mathbf{n}\rangle_T = \frac{1}{T} \int_0^T \mathbf{n}(t)dt,
\end{equation}
\begin{equation}
    \underline{\mathbf{n}} = n_x - in_y.
\end{equation}
Here, $n_x$ and $n_y$ are the projections of $\mathbf{n}$ on the $x$ and $y$ directions. Note that $\underline{\mathbf{n}}$ is a complex number, and that the real and imaginary parts of $\overline{\underline{\mathbf{n}}} = n_x+in_y$ correspond to the $x$ and $y$ components of $\mathbf{n}$. We note the following properties for integrable $T$-periodic functions $n_1(t)$ and $n_2(t)$ such that $\langle n_1\rangle_T = \langle n_2\rangle_T = 0$:
\begin{equation}
    \langle n_1 \tilde{n}_1\rangle_T = 0,
    \label{prop1}
\end{equation}
\begin{equation}
    \langle n_1 \tilde{n}_2\rangle_T = - \langle \tilde{n}_1 n_2 \rangle_T. 
     \label{prop2}
\end{equation}
In particular, because $\delta_1$ and $\delta_2$ are $T$-periodic, $\dot \delta_1$ and $\dot \delta_2$ are also $T$-periodic and $\langle \dot \delta_1\rangle_T = \langle \dot \delta_2\rangle_T = 0$. It follows from this that $\langle \Sigma\rangle_T = \langle \Gamma\rangle_T = 0$, and therefore $\langle U\rangle_T = \langle V\rangle_T = 0$, and $ \langle \mathbf{u}_E\rangle_T = 0$. Hence, properties (\ref{prop1}) and (\ref{prop2}) hold for $\dot \delta_1$, $\dot \delta_2$, $\Sigma$, $\Gamma$, $U$, and $V$.
Another useful property is the following:
\begin{equation}
    \underline{(\mathbf{n}_1\cdot\nabla)\mathbf{n}_2} = \overline{\underline{\mathbf{n}_1}} \ \underline{\mathbf{n'}_2}.
    \label{prop3}
\end{equation}

\subsection{Computation of the drift over one period}

The drift of a particle with initial position $\mathbf{x}_0$ is the integral of the Lagrangian velocity $\mathbf{u}_L(\mathbf{x}_0,t)$  over an actuation period. $\mathbf{u}_L$ is given by:
\begin{equation}
    \mathbf{u}_L(\mathbf{x}_0,t) = \mathbf{u}_E\left(\mathbf{x}_0+\tilde{\mathbf{u}}_E,t\right).
\end{equation}
At first order, the drift given by the potential flow solution vanishes over a period $T$:
\begin{equation}
    \langle\mathbf{u}_L\rangle_T \approx \langle\mathbf{u}_E\rangle_T = 0.
\end{equation}
At second order, the drift is:
\begin{equation}
    \langle\mathbf{u}_L\rangle_T \approx \langle \tilde{\mathbf{u}}_E\cdot \nabla \mathbf{u}_E \rangle_T.
\end{equation}
Because $u$ and $v$ are linear combinations of $U$ and $V$, there exist vectors $\mathbf{a}$ and $\mathbf{b}$ such that:
\begin{equation}
    \mathbf{u}_E(\mathbf{x},t) = U(t) \mathbf{a}(\mathbf{x}) + V(t) \mathbf{b}(\mathbf{x}).
    \label{defab}
\end{equation}
The drift is therefore given by:
\begin{equation}
    \langle \tilde{\mathbf{u}}_E\cdot \nabla \mathbf{u}_E \rangle_T = \langle \tilde UU \rangle_T (\mathbf{a} \cdot \nabla)\mathbf{a} + \langle \tilde VV \rangle_T (\mathbf{b} \cdot \nabla)\mathbf{b} + \langle \tilde UV \rangle_T (\mathbf{a} \cdot \nabla)\mathbf{b} + \langle \tilde VU \rangle_T (\mathbf{b} \cdot \nabla)\mathbf{a}.
\end{equation}
Using properties (\ref{prop1}) and (\ref{prop2}), this reduces to:
\begin{equation}
     \langle \tilde{\mathbf{u}}_E\cdot \nabla \mathbf{u}_E \rangle_T = \langle \tilde UV \rangle_T \mathbf{K},
\end{equation}
where:
\begin{equation}
    \mathbf{K} = (\mathbf{a} \cdot \nabla)\mathbf{b} - (\mathbf{b} \cdot \nabla)\mathbf{a}.
\end{equation}
We first evaluate $\langle \tilde UV \rangle_T$. From their respective definitions and Eq.~(\ref{prop1}):
\begin{equation}
    \langle \tilde UV \rangle_T = \frac{\sin(kx_c)\cos(kx_c)}{k} H(y_c) \langle\left[\tilde\Sigma + \tilde \Gamma G(y_c) \right]\Gamma \rangle_T 
    = \frac{\sin(2kx_c)}{2k} H(y_c) \langle \tilde \Sigma \Gamma\rangle_T.
\end{equation}
Substituting $\Sigma$ and $\Gamma$ from their respective definitions:
\begin{equation}
    \langle \tilde UV \rangle_T = \frac{\sin(2kx_c)}{2k^3h_0^2L_y} \left[ 1-\frac{\cosh(ky_c)}{\cosh(kL_y/2)} \right] \langle \delta_1 \dot \delta_2 \rangle_T.
\end{equation}
We now compute $\mathbf{K}$. From the chosen flow profile (Eq.~(\ref{cylinderflow})) and the definitions of $\mathbf{a}$ and $\mathbf{b}$ (Eq.~(\ref{defab})):
\begin{equation}
    \underline{\mathbf{u}}_E = \overline{\mathcal{U}} - \mathcal{U} \frac{R^2}{z^2} = U(t) \underline{\mathbf{a}} + V(t)\underline{\mathbf{b}}.
\end{equation}
Because $\underline{\mathbf{u}}_E$ is linear in $U$ and $V$, we identify:
\begin{equation}
   \underline{\mathbf{a}} = 1 - \frac{R^2}{z^2},
\end{equation}
\begin{equation}
   \underline{\mathbf{b}} = -i\left(1+\frac{R^2}{z^2}\right).
\end{equation}
Then, we note using (\ref{prop3}) that:
\begin{equation}
    \underline{\mathbf{K}} = \overline{\underline{\mathbf{a}}} \ \underline{\mathbf{b'}} - \overline{\underline{\mathbf{b}}} \ \underline{\mathbf{a'}}.
\end{equation}
This can be easily computed using the previous result:
\begin{equation}
    \overline{\underline{\mathbf{a}}} \ \underline{\mathbf{b'}} = \frac{2iR^2}{z^3} - \frac{2iR^4}{z^3 \overline{z}^2},
\end{equation}
\begin{equation}
    \overline{\underline{\mathbf{b}}} \ \underline{\mathbf{a'}} = \frac{2iR^2}{z^3} + \frac{2iR^4}{z^3 \overline{z}^2},
\end{equation}
so that:
\begin{equation}
    \underline{\mathbf{K}} = -\frac{4iR^4}{z^3 \overline{z}^2},
\end{equation}
and
\begin{equation}
    \overline{\underline{\mathbf{K}}} = \frac{4iR^4}{z^2 \overline{z}^3}.
\end{equation}
Using the polar coordinates $(r,\theta)$ centered at the obstacle such that $z=re^{i\theta}$:
\begin{equation}
    \overline{\underline{\mathbf{K}}} = \frac{4R^4}{r^5} i e^{i\theta}.
\end{equation}
By identification of the real and imaginary parts of $\overline{\underline{\mathbf{K}}}$ we find the $x$ and $y$ components of $\mathbf{K}$. Therefore:
\begin{equation}
    \mathbf{K} = \frac{4R^4}{r^5}(-\sin\theta \ \mathbf{e}_x + \cos\theta \ \mathbf{e}_y) = \frac{4R^4}{r^5} \mathbf{e}_\theta,
\end{equation}
where $\mathbf{e}_x$, $\mathbf{e}_y$, and $\mathbf{e}_\theta$ are the unit vectors in the $x$, $y$, and $\theta$ directions.
The drift can then be written as:
\begin{equation}
    \langle \tilde{\mathbf{u}}_E\cdot \nabla \mathbf{u}_E \rangle_T = \frac{2R^4}{r^5} \frac{\sin(2kx_c)}{k^3h_0^2L_y} \left[ 1-\frac{\cosh(ky_c)}{\cosh(kL_y/2)} \right] \langle \delta_1 \dot \delta_2 \rangle_T \ \mathbf{e}_\theta.
\end{equation}
Finally, we can note that $\langle \delta_1 \dot \delta_2 \rangle_T$ is related to the signed area $\mathcal{A}$ enclosed by the actuation loop in the $(\delta_1,\delta_2)$ space. Indeed:
\begin{equation}
    \langle \delta_1 \dot \delta_2 \rangle_T = \frac{1}{T} \int_0^T \delta_1 \dot \delta_2 dt = \frac{1}{T} \oint\delta_1 d\delta_2=\frac{\mathcal{A}}{T}.
\end{equation}
This yields the final result:
\begin{equation}
    \langle \tilde{\mathbf{u}}_E\cdot \nabla \mathbf{u}_E \rangle_T = \frac{2R^4}{r^5} \frac{\sin(2kx_c)}{k^3h_0^2L_y} \left[ 1-\frac{\cosh(ky_c)}{\cosh(kL_y/2)} \right] \frac{\mathcal{A}}{T} \ \mathbf{e}_\theta.
\end{equation}
In the case of sinusoidal actuation of equal amplitude, $\mathcal{A}$ is the area of a Lissajous ellipse. Indeed, if the ceiling deformation is $\delta_i=\frac{\delta_0}{2} \left(1+\sin\left(\frac{2\pi t}{T} + \{0, \varphi\}\right)\right)$, then:
\begin{equation}
    \mathcal{A} = \frac{\pi \delta_0^2}{2T} \int_0^T \sin\left(\frac{2\pi t}{T}\right) \cos\left(\frac{2\pi t}{T} + \varphi\right) dt = -\frac{\pi }{4} \delta_0^2 \sin \varphi,
\end{equation}
and:
\begin{equation}
    \langle \tilde{\mathbf{u}}_E\cdot \nabla \mathbf{u}_E \rangle_T = \frac{ R^4}{4r^5} \frac{\sin(2kx_c)}{k^3h_0^2L_y} \left[ \frac{\cosh(ky_c)}{\cosh(kL_y/2)} -1\right] \omega \delta_0^2 \sin \varphi \ \mathbf{e}_\theta.
\end{equation} 

%% file: biblio.bib
@article{Kriezis1992PI,
  title = {Eddy Currents: Theory and Applications},
  shorttitle = {Eddy Currents},
  author = {Kriezis, E. E. and Tsiboukis, T. D. and Panas, S. M. and Tegopoulos, J. A.},
  year = 1992,
  month = oct,
  journal = {Proceedings of the IEEE},
  volume = {80},
  number = {10},
  pages = {1559--1589},
  issn = {1558-2256},
  doi = {10.1109/5.168666},
  urldate = {2026-09-24}
}

@article{Dabiri2005JEB,
  title = {Flow Patterns Generated by Oblate Medusan Jellyfish: Field Measurements and Laboratory Analyses},
  shorttitle = {Flow Patterns Generated by Oblate Medusan Jellyfish},
  author = {Dabiri, John O. and Colin, Sean P. and Costello, John H. and Gharib, Morteza},
  year = 2005,
  month = apr,
  journal = {Journal of Experimental Biology},
  volume = {208},
  number = {7},
  pages = {1257--1265},
  issn = {0022-0949},
  doi = {10.1242/jeb.01519},
  urldate = {2026-09-24},
}

@article{Sommeria1988N,
  title = {Laboratory Simulation of {{Jupiter}}'s {{Great Red Spot}}},
  author = {Sommeria, J{\"o}el and Meyers, Steven D. and Swinney, Harry L.},
  year = 1988,
  month = feb,
  journal = {Nature},
  volume = {331},
  number = {6158},
  pages = {689--693},
  publisher = {Nature Publishing Group},
  issn = {1476-4687},
  doi = {10.1038/331689a0},
  urldate = {2026-09-24},
  copyright = {1988 Springer Nature Limited},
  langid = {english},
}

@article{Moffatt1964JFM,
  title = {Viscous and Resistive Eddies near a Sharp Corner},
  author = {Moffatt, H. K.},
  year = 1964,
  month = jan,
  journal = {Journal of Fluid Mechanics},
  volume = {18},
  number = {1},
  pages = {1--18},
  issn = {1469-7645, 0022-1120},
  doi = {10.1017/S0022112064000015},
  urldate = {2026-09-24},
}

@article{Jakiela2012PRL,
  title = {Discontinuous {{Transition}} in a {{Laminar Fluid Flow}}: {{A Change}} of {{Flow Topology}} inside a {{Droplet Moving}} in a {{Micron-Size Channel}}},
  shorttitle = {Discontinuous {{Transition}} in a {{Laminar Fluid Flow}}},
  author = {Jakiela, Slawomir and Korczyk, Piotr M. and Makulska, Sylwia and Cybulski, Olgierd and Garstecki, Piotr},
  year = 2012,
  month = mar,
  journal = {Physical Review Letters},
  volume = {108},
  number = {13},
  pages = {134501},
  issn = {0031-9007, 1079-7114},
  doi = {10.1103/PhysRevLett.108.134501},
  urldate = {2026-09-24},
  copyright = {http://link.aps.org/licenses/aps-default-license},
}

@book{Panton2013,
  title={{Incompressible Flow}},
  author={Panton, Ronald L},
  year={2013},
  edition = {4},
  publisher={John Wiley \& Sons}
}

@article{Gallaire2014PF,
  title = {Marangoni Induced Force on a Drop in a {{Hele Shaw}} Cell},
  author = {Gallaire, Fran{\c c}ois and Meliga, Philippe and Laure, Patrice and Baroud, Charles N.},
  year = 2014,
  month = jun,
  journal = {Physics of Fluids},
  volume = {26},
  number = {6},
  pages = {062105},
  issn = {1070-6631, 1089-7666},
  doi = {10.1063/1.4878095},
  urldate = {2026-09-22},
  langid = {english},
}

@article{McKee2025PRF,
  title = {Potential Flows with Electromagnetically Induced Circulation in a {{Hele-Shaw}} Cell},
  author = {McKee, Kyle I. and Bush, John W. M.},
  year = 2025,
  month = may,
  journal = {Physical Review Fluids},
  volume = {10},
  number = {5},
  pages = {054103},
  publisher = {American Physical Society},
  doi = {10.1103/PhysRevFluids.10.054103}
}

@article{Rallabandi2014JFM,
  title = {Two-Dimensional Streaming Flows Driven by Sessile Semicylindrical Microbubbles},
  author = {Rallabandi, Bhargav and Wang, Cheng and Hilgenfeldt, Sascha},
  year = 2014,
  month = jan,
  journal = {Journal of Fluid Mechanics},
  volume = {739},
  pages = {57--71},
  issn = {0022-1120, 1469-7645},
  doi = {10.1017/jfm.2013.616},
  copyright = {https://www.cambridge.org/core/terms}
}

@article{Zhang2023JFM,
  title = {Three-Dimensional Streaming around an Obstacle in a {{Hele-Shaw}} Cell},
  author = {Zhang, Xirui and Rallabandi, Bhargav},
  year = 2023,
  month = apr,
  journal = {Journal of Fluid Mechanics},
  volume = {961},
  pages = {A35},
  issn = {0022-1120, 1469-7645},
  doi = {10.1017/jfm.2023.276}
}

@article{Loutherback2009PRL,
  title = {Deterministic {{Microfluidic Ratchet}}},
  author = {Loutherback, Kevin and Puchalla, Jason and Austin, Robert H. and Sturm, James C.},
  year = 2009,
  month = jan,
  journal = {Physical Review Letters},
  volume = {102},
  number = {4},
  pages = {045301},
  issn = {0031-9007, 1079-7114},
  doi = {10.1103/PhysRevLett.102.045301},
  urldate = {2026-09-22},
  copyright = {http://link.aps.org/licenses/aps-default-license},
}

@article{Kurzthaler2024JFM,
  title = {Surface Corrugations Induce Helical Near-Surface Flows and Transport in Microfluidic Channels},
  author = {Kurzthaler, Christina and Chase, Danielle L. and Stone, Howard A.},
  year = 2024,
  month = mar,
  journal = {Journal of Fluid Mechanics},
  volume = {982},
  pages = {A31},
  issn = {0022-1120, 1469-7645},
  doi = {10.1017/jfm.2024.106},
  urldate = {2026-09-24},
  langid = {english},
}

@article{Norden2002APL,
  title = {Ratchet Device with Broken Friction Symmetry},
  author = {Nord{\'e}n, Bengt and Zolotaryuk, Yaroslav and Christiansen, Peter L. and Zolotaryuk, Alexander V.},
  year = 2002,
  month = apr,
  journal = {Applied Physics Letters},
  volume = {80},
  number = {14},
  pages = {2601--2603},
  issn = {0003-6951},
  doi = {10.1063/1.1468900}
}

@article{Stokes1847TCPS,
  title = {On the Theory of Oscillatory Waves},
  author = {Stokes, George Gabriel},
  year = 1847,
  journal = {Transactions of the Cambridge Philosophical Society},
  volume = {8},
  pages = {441--455}
}

@article{Longuet-Higgins1970JFM,
  title = {Steady Currents Induced by Oscillations Round Islands},
  author = {{Longuet-Higgins}, M. S.},
  year = 1970,
  month = jul,
  journal = {Journal of Fluid Mechanics},
  volume = {42},
  number = {4},
  pages = {701--720},
  issn = {1469-7645, 0022-1120},
  doi = {10.1017/S0022112070001568},
  urldate = {2026-09-22},
}

@article{vandenBremer2018PTRSA,
  title = {Stokes Drift},
  author = {{van den Bremer}, T. S. and Breivik, {\O}.},
  year = 2018,
  month = jan,
  journal = {Philosophical Transactions of the Royal Society A: Mathematical, Physical and Engineering Sciences},
  volume = {376},
  number = {2111},
  pages = {20170104},
  issn = {1364-503X},
  doi = {10.1098/rsta.2017.0104},
  urldate = {2026-09-22},
}

@misc{Gebhard2026,
  title = {Designing Single-Layer {{PDMS}} Devices for Micron to Millimeter-Scale Deformations},
  author = {Gebhard, Leon Valentin and Avaro, Alexandre S. and Amselem, Gabriel and Baroud, Charles N.},
  year = 2026,
  month = may,
  number = {arXiv:2605.17402},
  eprint = {2605.17402},
  primaryclass = {physics.flu-dyn},
  publisher = {arXiv},
  doi = {10.48550/arXiv.2605.17402},
  urldate = {2026-06-01},
  archiveprefix = {arXiv}
}

@article{Jain2024B,
  title = {Using a Micro-Device with a Deformable Ceiling to Probe Stiffness Heterogeneities within {{3D}} Cell Aggregates},
  author = {Jain, Shreyansh and Belkadi, Hiba and Michaut, Arthur and Sart, S{\'e}bastien and Gros, J{\'e}r{\^o}me and Genet, Martin and Baroud, Charles N},
  year = 2024,
  month = apr,
  journal = {Biofabrication},
  volume = {16},
  number = {3},
  pages = {035010},
  publisher = {IOP Publishing},
  issn = {1758-5090},
  doi = {10.1088/1758-5090/ad30c7},
  urldate = {2026-06-01},
  langid = {english},
}

@article{Dangla2011PRL,
  title = {Trapping {{Microfluidic Drops}} in {{Wells}} of {{Surface Energy}}},
  author = {Dangla, R{\'e}mi and Lee, Sungyon and Baroud, Charles N.},
  year = 2011,
  month = sep,
  journal = {Physical Review Letters},
  volume = {107},
  number = {12},
  pages = {124501},
  publisher = {American Physical Society},
  doi = {10.1103/PhysRevLett.107.124501},
  urldate = {2026-09-24},
}

@article{Sequeira2023JCED,
  title = {Viscosity and {{Density Measurements}} of {{Poly}}(Ethyleneglycol) 200 and {{Poly}}(Ethyleneglycol) 600 at {{High Pressures}}},
  author = {Sequeira, Maria C. M. and Avelino, Helena M. N. T. and Caetano, Fernando J. P. and Fareleira, Jo{\~a}o M. N. A.},
  year = 2023,
  month = jan,
  journal = {Journal of Chemical \& Engineering Data},
  volume = {68},
  number = {1},
  pages = {64--72},
  publisher = {American Chemical Society},
  issn = {0021-9568},
  doi = {10.1021/acs.jced.2c00578}
}

@article{Thielicke2014JORS,
  title = {{{PIVlab}} -- {{Towards User-friendly}}, {{Affordable}} and {{Accurate Digital Particle Image Velocimetry}} in {{MATLAB}}},
  author = {Thielicke, William and Stamhuis, Eize J.},
  year = 2014,
  month = oct,
  journal = {Journal of Open Research Software},
  volume = {2},
  number = {1},
  issn = {2049-9647},
  pages = {e30},
  doi = {10.5334/jors.bl}
}

@book{Milne-Thomson1960,
  title = {Theoretical {{Hydrodynamics}}},
  author = {{Milne-Thomson}, Louis Melville},
  year = 1960,
  publisher = {Macmillan},
  langid = {english},
  edition = {4},
}

@article{Lester2009PRE,
  title = {Lagrangian Topology of a Periodically Reoriented Potential Flow: {{Symmetry}}, Optimization, and Mixing},
  shorttitle = {Lagrangian Topology of a Periodically Reoriented Potential Flow},
  author = {Lester, D. R. and Metcalfe, G. and Trefry, M. G. and Ord, A. and Hobbs, B. and Rudman, M.},
  year = 2009,
  month = sep,
  journal = {Physical Review E},
  volume = {80},
  number = {3},
  pages = {036208},
  issn = {1539-3755, 1550-2376},
  doi = {10.1103/PhysRevE.80.036208},
  urldate = {2026-09-01},
  copyright = {http://link.aps.org/licenses/aps-default-license},
  langid = {english},
}

@article{Purcell1977AJP,
  title = {Life at Low {{Reynolds}} Number},
  author = {Purcell, E. M.},
  year = 1977,
  month = jan,
  journal = {American Journal of Physics},
  volume = {45},
  number = {1},
  pages = {3--11},
  issn = {0002-9505},
  doi = {10.1119/1.10903}
}

@article{Shapere1989JFM,
  title = {Geometry of Self-Propulsion at Low {{Reynolds}} Number},
  author = {Shapere, Alfred and Wilczek, Frank},
  year = 1989,
  month = jan,
  journal = {Journal of Fluid Mechanics},
  volume = {198},
  pages = {557--585},
  issn = {0022-1120, 1469-7645},
  doi = {10.1017/S002211208900025X},
  copyright = {https://www.cambridge.org/core/terms}
}

@article{Amselem2023PRA,
  title = {Valveless {{Pumping}} at {{Low Reynolds Numbers}}},
  author = {Amselem, Gabriel and Clanet, Christophe and Benzaquen, Michael},
  year = 2023,
  month = feb,
  journal = {Physical Review Applied},
  volume = {19},
  number = {2},
  pages = {024017},
  publisher = {American Physical Society},
  doi = {10.1103/PhysRevApplied.19.024017},
  urldate = {2026-06-01},
}

@article{Box2020PNAS,
  title = {Flow-Induced Choking of a Compliant {{Hele-Shaw}} Cell},
  author = {Box, Finn and Peng, Gunnar G. and {Pihler-Puzovi{\'c}}, Draga and Juel, Anne},
  year = 2020,
  month = dec,
  journal = {Proceedings of the National Academy of Sciences},
  volume = {117},
  number = {48},
  pages = {30228--30233},
  publisher = {Proceedings of the National Academy of Sciences},
  doi = {10.1073/pnas.2008273117},
  urldate = {2026-09-24},
  langid = {english},
}

@misc{SM,
 title={See {Supplemental Material} at [{URL}] for the derivation of the flow potential and drift, {Fig. S1}, and {Movies S1, S2, and S3}.} }
